\documentstyle[12pt,epsfig]{article}
\newlength{\dinwidth} 
\newlength{\dinmargin}
\begin{document}
%
\def\rsep{R_{\rm sep}}
\def\cts{\cos\theta^{\ast}} 
\def\xgo{x_\gamma^{OBS}}
\def\xpo{x_p^{OBS}} 
\def\xg{x_\gamma} 
\def\ETJ{E_T^{jet}}
\def\ETCJ{E_T^{CALJet}}
\def\ETTJ{E_T^{TLTJet}}
\def\ETJM{E_T^{min}} 
\def\ETAJ{\eta^{jet}} 
\def\PHIJ{\phi^{jet}} 
\def\ETACJ{\eta^{CALJet}} 
\def\ETATJ{\eta^{TLTJet}} 
\def\ETAB{\bar{\eta}}
\def\EEP{E^\prime_{e}} 
\def\TEP{\theta^\prime_{e}} 
\def\MJJ{M_{jj}}
\def\DETA{\Delta\eta} 
\def\ptmin{\hat{p}_T^{\rm min}}

\def\3{\ss}

\begin{titlepage}

\title{Dijet Cross Sections in Photoproduction at HERA}

\author{ZEUS Collaboration}

\date{}

\maketitle

\vspace{-6.5 cm}
\begin{flushright}
DESY 97-196
\end{flushright}
\vspace{10.5 cm}

\begin{abstract}   
Dijet cross sections are presented using photoproduction data obtained
with the ZEUS detector during 1994. These measurements represent an
extension of previous results, as the higher statistics allow cross
sections to be measured at higher jet transverse energy ($\ETJ$).
Jets are identified in the hadronic final state using three different
algorithms, and the cross sections compared to complete
next-to-leading order QCD calculations.  Agreement with these
calculations is seen for the pseudorapidity dependence of the direct
photon events with $\ETJ > 6$~GeV and of the resolved photon events
with $\ETJ > 11$ GeV. Calculated cross sections for resolved photon
processes with 6~GeV $< \ETJ < 11$~GeV lie below the data.
\end{abstract} 


\setcounter{page}{0}
\thispagestyle{empty}

\newpage

\end{titlepage}

\newpage

%
%
%
%
                                                   %
\begin{center}                                                                                     
{                      \Large  The ZEUS Collaboration              }                               
\end{center}                                                                                       
  J.~Breitweg,                                                                                     
  M.~Derrick,                                                                                      
  D.~Krakauer,                                                                                     
  S.~Magill,                                                                                       
  D.~Mikunas,                                                                                      
  B.~Musgrave,                                                                                     
  J.~Repond,                                                                                       
  R.~Stanek,                                                                                       
  R.L.~Talaga,                                                                                     
  R.~Yoshida,                                                                                      
  H.~Zhang  \\                                                                                     
 {\it Argonne National Laboratory, Argonne, IL, USA}~$^{p}$
% ...
\newpage                                                                                           
% ...
$^{  23}$ now at ATLAS Collaboration, Univ. of Munich \\                                           
% ...
$^{  32}$ partially supported by the Foundation for German-Russian Collaboration                   
DFG-RFBR \\ \hspace*{3.5mm} (grant no. 436 RUS 113/248/3 and no. 436 RUS 113/248/2)\\              
% ...
                                                           %
                                                           %
\newpage   
                                                           %
                                                           %
\begin{tabular}[h]{rp{14cm}}                                                                       
$^{a}$ &  supported by the Natural Sciences and Engineering Research                               
          Council of Canada (NSERC)  \\                                                            
$^{b}$ &  supported by the FCAR of Qu\'ebec, Canada  \\                                            
$^{c}$ &  supported by the German Federal Ministry for Education and                               
          Science, Research and Technology (BMBF), under contract                                  
          numbers 057BN19P, 057FR19P, 057HH19P, 057HH29P, 057SI75I \\                              
$^{d}$ &  supported by the MINERVA Gesellschaft f\"ur Forschung GmbH,                              
          the German Israeli Foundation, and the U.S.-Israel Binational                            
          Science Foundation \\                                                                    
$^{e}$ &  supported by the German Israeli Foundation, and                                          
          by the Israel Science Foundation                                                         
  \\                                                                                               
$^{f}$ &  supported by the Italian National Institute for Nuclear Physics                          
          (INFN) \\                                                                                
$^{g}$ &  supported by the Japanese Ministry of Education, Science and                             
          Culture (the Monbusho) and its grants for Scientific Research \\                         
$^{h}$ &  supported by the Korean Ministry of Education and Korea Science                          
          and Engineering Foundation  \\                                                           
$^{i}$ &  supported by the Netherlands Foundation for Research on                                  
          Matter (FOM) \\                                                                          
$^{j}$ &  supported by the Polish State Committee for Scientific                                   
          Research, grant No.~115/E-343/SPUB/P03/002/97, 2P03B10512,                               
          2P03B10612, 2P03B14212, 2P03B10412 \\                                                    
$^{k}$ &  supported by the Polish State Committee for Scientific                                   
          Research (grant No. 2P03B08308) and Foundation for                                       
          Polish-German Collaboration  \\                                                          
$^{l}$ &  partially supported by the German Federal Ministry for                                   
          Education and Science, Research and Technology (BMBF)  \\                                
$^{m}$ &  supported by the Fund for Fundamental Research of Russian Ministry                       
          for Science and Edu\-cation and by the German Federal Ministry for                       
          Education and Science, Research and Technology (BMBF) \\                                 
$^{n}$ &  supported by the Spanish Ministry of Education                                           
          and Science through funds provided by CICYT \\                                           
$^{o}$ &  supported by the Particle Physics and                                                    
          Astronomy Research Council \\                                                            
$^{p}$ &  supported by the US Department of Energy \\                                              
$^{q}$ &  supported by the US National Science Foundation \\                                       
\end{tabular}                                                                                      
                                                           %
                                                           %

\newpage
\setcounter{page}{1}
\parindent 5mm

\section{Introduction} 
High energy collisions between photons and protons can produce jets in
the final state.  In leading order quantum chromodynamics (LO QCD),
two types of processes lead to the photoproduction of jets. In direct
processes the photon participates in the hard scatter via either
boson-gluon fusion or QCD Compton scattering. In resolved processes
the photon acts as a source of quarks and gluons, and only a fraction
of its momentum participates in the hard scatter. This separation
between direct and resolved photoproduction is only well defined in
this way at leading order. To make a measurement which can be compared
to calculations at any order, the variable $\xgo$ is used to separate
these two types of event~\cite{samerapdijet}.  The variable $\xgo$ is
the fraction of the photon's momentum contributing to the production
of the two highest transverse energy ($\ETJ$) jets. It is defined for
the photoproduction of jets in positron-proton scattering as:
\begin{equation}
  \xgo = \frac{E_{T}^{jet1}e^{-\eta^{jet1}} + E_{T}^{jet2}e^{-\eta^{jet2}}}
{2yE_e}
\end{equation}
where $E_e$ is the initial positron energy and $\ETAJ$ is the jet
pseudorapidity\footnote{The pseudorapidity is defined as $\eta =
-$ln$(\tan\frac{\theta}{2}$) where $\theta$ is the polar angle with
respect to the $Z$ axis, which in the ZEUS coordinate system is
defined to be the proton beam direction.}. The inelasticity $y$ is defined
in the ZEUS frame as $y = 1 - \frac{\EEP}{2 E_e} (1-\cos{\TEP})$ where
$\EEP$ and $\TEP$ are the energy and polar angle of the outgoing
positron. In a leading order calculation, direct processes have
$\xgo=1$ since all the photon's momentum participates in the
production of the high transverse energy jets, while resolved
processes have $\xgo < 1$ since part of the photon's momentum goes
into the photon remnant.  Throughout the following, in both the data
and the calculations, direct and resolved samples are defined in terms
of a cut on $\xgo$ rather than in terms of the LO diagrams.
In a recent analysis by the H1 collaboration, a similar variable was used
to determine an effective parton distribution in the photon~\cite{H1eff}.

In a previous analysis~\cite{samerapdijet} dijet cross sections were
measured using 1993 ZEUS data in the kinematic regime where the
difference between the pseudorapidities of the two jets is small
($|\Delta \eta|= |\eta^{jet1} - \eta^{jet2}| < 0.5$).  This condition
constrains $\theta^{\ast}$, the angle between the jet-jet axis and the
beam axis in the dijet centre of mass system, to be close to $90^o$.
The cross section as a function of $\ETAB = (\eta^{jet1} +
\eta^{jet2})/2$ then has maximal sensitivity to the parton
distributions in both the photon and proton~\cite{FR}. 
In \cite{samerapdijet}, the comparison between data and Monte Carlo
(MC) simulations based on the LO direct and resolved processes showed
that the jet profiles, as described by the transverse energy flow
around the jet axis, are poorly reproduced for jets with low $\ETJ$
produced in the forward (proton) direction. In the present analysis a
comparison will be made with MC simulations which include multiparton
interactions, and an improved description of the data is obtained.

To compare data and theoretical cross sections based on
next-to-leading order (NLO) QCD calculations, it is essential that
similar jet definitions be employed for both the measurement and
calculations.
The dijet cross sections as a function of $\ETJ$ and $\ETAB$, for low
and high $\xgo$, are measured in the hadronic final state using
various jet definitions, including the $k_T$ algorithm. The resulting
cross sections are compared to NLO QCD calculations at the parton
level. The uncertainties due to hadronization effects are not yet
theoretically estimated and are not considered in the
comparison. After a brief description of the experimental setup, a
discussion of the issues involved in the various jet definitions in
both theory and experiment is presented, followed by our results and
conclusions.

\section{Experimental Setup}
In 1994 HERA provided 27.5~GeV positrons and 820~GeV protons colliding
in 153 bunches. Additional unpaired positron and proton bunches
circulated to allow monitoring of the background from beam-gas
interactions. Events from empty beam crossings (that is bunches
containing neither positrons nor protons) were used to estimate the
background from cosmic rays.  The total integrated luminosity used in
this analysis is $2.70$~pb$^{-1}$ with an uncertainty of $\pm$1.5\%.

Details of the ZEUS detector have been described
elsewhere~\cite{ZEUS}. The primary components used in this analysis
are the central tracking system and the calorimeter.  The central
tracking system consists of a vertex detector~\cite{VXD} and a central
tracking detector~\cite{CTD} enclosed in a 1.43~T solenoidal magnetic
field.  The uranium and scintillator calorimeter~\cite{CAL} covers
99.7\% of the total solid angle and is subdivided into three parts:
forward (FCAL) covering $4.3 > \eta > 1.1$, barrel (BCAL) covering the
central region $1.1 > \eta > -0.75$ and rear (RCAL) covering the
backward region $-0.75 > \eta > -3.8$, for a collision at the nominal
interaction point.  Each calorimeter part consists of an
electromagnetic section followed by an hadronic section. The cells of
these sections have inner face sizes of $5 \times 20$
cm$^2$ ($10 \times 20$ cm$^2$ in the rear calorimeter) and $20 \times
20$ cm$^2$, respectively.  A lead and scintillator calorimeter is used
to measure the luminosity via the the detection of photons from the
positron-proton brems\-strahlung process.  This calorimeter is
installed 107~m along the HERA tunnel from the interaction point in
the positron direction and subtends a small angle at the interaction
vertex~\cite{LUMI}. A fraction of the positrons scattered through
small angles are detected in a similar lead and scintillator
calorimeter positioned at $Z = -35$~m.

\section{Jet Algorithms}

Most of the previous measurements of jet cross sections at
hadron-hadron colliders and in photoproduction at HERA have used some
variation of a cone-based jet algorithm.  In these algorithms,
according to the standardisation of cone jet algorithms at the
Snowmass meeting in 1990 \cite{Hut92}, jets consist of calorimeter
cells (or, in a theoretical description, partons) $i$ with a distance
\begin{equation}
\label{snow-eqn}
R_i = \sqrt{(\eta_i-\ETAJ)^2+(\phi_i-\PHIJ)^2} \leq R \label{eq1}
\end{equation} 
from the jet centre.  Here $\phi_i$ and $\eta_i$ are the azimuthal
angle and pseudorapidity of the cell (or parton), and $R$ is the jet cone
radius. In this analysis, the geometric centre of the cell is used
to define the position.  The parameters for the jet are calculated
as:
\begin{eqnarray}
                   \ETJ &=& \sum_i E_{T_i}\nonumber\\
\label{jetpar_eqn} \eta^{jet} &=& \frac{1}{\ETJ}\sum_i E_{T_i} \eta_i\\
                   \phi^{jet} &=& \frac{1}{\ETJ{}}\sum_i E_{T_i}
\phi_i\nonumber
\end{eqnarray}
in which the sums run over all calorimeter cells (or partons)
belonging to the jet.  Different approaches are possible to the
choice of the `seed' with which to begin jet finding, and to how and
when overlapping jets are merged. The approach is not fixed
by the Snowmass convention. We use two different cone algorithms 
to determine dijet cross sections in
photoproduction. The jet cone radius $R=1$ is chosen for both
algorithms. We also use a cluster algorithm, which does not suffer
from these ambiguities. A further advantage of the cluster algorithm
is that it is infrared safe to all orders, which is not always the
case for cone algorithms~\cite{safe}.
In the following the three algorithms will be described
in detail considering as an example the case of calorimeter cells. Identical
algorithms are used in this analysis to define jets in the hadronic 
final state starting from the final state particles.

In the first cone algorithm (EUCELL) a window in the $\eta-\phi$ space 
of the calorimeter cells is
moved around to find those positions where the $E_T$ in the window is
$> 1$~GeV to use as seeds.  
The jet quantities are initially calculated using the cells in a cone centred on the seed.
Equations (\ref{snow-eqn})
and~(\ref{jetpar_eqn}) are then applied to choose the cells belonging
to the jets and to update the jet quantities in an iterative procedure
until a stable jet is found.
Only the highest transverse energy jet is
accepted, the cells within the jet are removed, and the whole process
is repeated. In this way EUCELL produces no overlapping jets.

The second cone algorithm (PUCELL) was adapted from the algorithm used
by CDF~\cite{cdf} and determines seeds by finding the single
calorimeter cell of
highest transverse energy and placing a cone around it. All the cells
within the selected cone are assigned to this seed and excluded from
the search for further seeds, which is then continued.  
The jet quantities are initially calculated using the cells in the seed
and equations (\ref{snow-eqn}) and~(\ref{jetpar_eqn}) are then applied
iteratively as for EUCELL until a stable jet is found.
At this stage
all jets are provisionally accepted.  Thus it may happen that two
stable jets overlap.  If the overlapping transverse energy amounts to
more than 75\% of the smallest jet, they are merged, otherwise the
overlapping energy is split such that cells are associated with the
closest jet.

In the cluster algorithm KTCLUS~\cite{Cat93,Ell93} the quantity
\begin{equation}
d_{i,j}=\left((\eta_i-\eta_j)^2+(\phi_i-\phi_j)^2\right) 
{\rm min}(E_{T_i},E_{T_j})^2
\end{equation}
is calculated for each pair of objects (where the initial objects are
the calorimeter cells), and for
each individual object:
\begin{equation}
d_i=E^2_{T_i}.
\end{equation}
If, of all the numbers [$d_{i,j},d_i$], $d_{k,l}$ is the smallest then
objects $k$ and $l$ are combined into a single new object.  If however
$d_k$ is the smallest, then object $k$ is a jet and is removed from
the sample.  This is repeated until all objects are assigned to jets.
As with the cone algorithms, eq.~(\ref{jetpar_eqn}) is used to
determine the parameters of the jets.  It is also used to determine
the parameters of the intermediate objects.  

Equations~(\ref{snow-eqn}) and~(\ref{jetpar_eqn}) imply that in a NLO
calculation, two partons must be a distance
\begin{equation} 
R_{ij} =\sqrt{(\eta_i-\eta_j)^2+(\phi_i-\phi_j)^2} \leq
\frac{E_{T_i}+E_{T_j}} {\max (E_{T_i},E_{T_j})} R \label{eq2}
\label{snow-parton-eqn}
\end{equation}
from each other to be combined, where $E_{T_k}$ is the transverse
energy of parton $k$. This implies that if two partons have
approximately equal transverse energy they may be separated from each
other by as much as $2R$ and still satisfy
eq.(\ref{snow-eqn}). However, as parton $j$ does not then lie inside a
cone of radius $R$ around parton $i$ and vice versa, one might with
some justification also count the two partons separately. If one
wishes to compare theory with measurement it is necessary to match the
theoretical treatment of such cases to the operation of the jet finding
and jet merging criteria used experimentally. This is done by
introducing an additional parameter, $R_{\rm sep}$, to the theory to
restrict the maximum separation between two partons in a single jet
\cite{Ell92}. Eq.~(\ref{snow-parton-eqn}) then becomes
\begin{equation} 
R_{ij} \leq \min\left[\frac{E_{T_i}+E_{T_j}}{\max(E_{T_i},E_{T_j})}R, R_{\rm
sep}\right].  
\end{equation} 
The valid range of $R_{\rm sep}$ is between $1R$ and $2R$. For a NLO
three parton final state, it is found that $R_{\rm sep} = (1.5-2.0)
\cdot R$ corresponds to EUCELL, $R_{\rm sep} = 1\cdot R$ to PUCELL,
and $R = R_{\rm sep} = 1$ to KTCLUS~\cite{bfkk}. In this paper, all
three jet definitions will be used for a comparison of the resulting
dijet cross sections.  An alternative approach would be to treat
$\rsep$ as a parameter, and tune it in order to take into account
possible theoretical uncertainties such as higher order
contributions. However, in the present analysis this approach has not
been followed and $\rsep$ is fixed by the functionality of each jet
algorithm.

\section{Event Selection}
\label{sec-ev-sel}
The ZEUS detector uses a three-level trigger system.  The first level
selects events used in this analysis with a coincidence of a regional
or transverse energy sum in the calorimeter, and at least one track
from the interaction point measured in the central tracking chamber.
At the second level, at least 8~GeV total transverse energy, excluding
the eight calorimeter towers immediately surrounding the forward
beampipe, is required, and cuts on calorimeter energies and timing are
used to suppress events caused by interactions between the proton beam
and residual gas in the beam pipe~\cite{F2}.  At the third level, a
cone algorithm uses the calorimeter cell energies
and positions to identify jets. Events are required to have at least
two jets, each of which has $\ETTJ > 3.5$~GeV and $\ETATJ < 2.0$ or
$\ETTJ > 4.0$~GeV and $2.0 < \ETATJ < 2.5$.  Additional tracking cuts
were made to reject proton beam-gas interactions and cosmic ray
events.

Further cuts are applied offline. Charged current deep inelastic
scattering events are rejected by a cut on the missing transverse
momentum measured in the calorimeter. To reject remaining beam-gas and
cosmic ray background events, tighter cuts using the final $Z-$vertex
position, other tracking information and timing information are
applied.  Two additional cuts are made~\cite{ZEUSdir+inc}, based upon
two different measurements of $y$:
\begin{enumerate}
\item 
Events with a positron candidate in the uranium calorimeter 
are removed if $y_e  < 0.7$, where $y_e$ is the value of $y$ 
as measured assuming the positron candidate is the scattered 
positron.
\item 
A cut is made on the Jacquet-Blondel measurement of $y$~\cite{YJB},
$y_{JB} = \sum_i (E_i - E_{zi}) /2E_e$, where $E_{zi} = E_i
\cos\theta_i$, and $E_i$ is the energy deposited in the calorimeter
cell $i$ which has a polar angle $\theta_i$ with respect to the
measured $Z$-vertex of the event.  The sum runs over all calorimeter
cells.  For any event where the scattered positron entered the uranium
calorimeter and either is not identified or gives $y_e$ above $0.7$,
the value of $y_{JB}$ will be near to one.  Proton beam-gas events
will have low values of $y_{JB}$.  To further reduce the contamination
from both these sources, it is required that $0.15 < y_{JB} < 0.7$.
This range corresponds approximately to the true $y$ range of $0.2 < y
< 0.8$.

\end{enumerate}
These cuts restrict the range of the photon virtuality to less than
$\sim4$~GeV$^{2}$, with a median of around $10^{-3}$~GeV$^2$, which
excludes deep inelastic scattering (DIS) events.

To select dijet candidates with a particular jet algorithm, the
algorithm is applied to the calorimeter cells. In each case, the jet
transverse energy measured in the ZEUS detector is corrected as a
function of $\ETACJ$ and $\ETCJ$. The variable $\ETCJ$ is used to
denote the transverse energy of a jet before correction for the
effects of inactive material. This correction is derived from the MC
events described in the next section by comparing the true transverse
energy of the jet, found by applying the algorithm to the final state
particles, to the (lower) transverse energy measured in the
calorimeter simulation, obtaining the average shift between the two
transverse energies for each jet algorithm.  The average shift in jet
energies is around $17\%$ for all three jet algorithms, and varies
between $10\%$ and $25\%$ depending upon $\eta^{jet}$. The largest
shifts occur at the boundaries between the FCAL and BCAL and between
the BCAL and RCAL.  No correction was applied to $\ETACJ$ since, from
MC, the average shift in $\eta$ between the particle and detector jets
is less than $\pm$0.05 for all $\eta$ values in the range used for the
cross section measurements.

The description of the calorimeter response to particles and jets in
the MC has been tuned using several methods~\cite{escale,theses},
including (i) the comparison of charged track momenta with calorimeter
energy measurements, (ii) comparison of jet and positron variables in
DIS events and (iii) the comparison of the measurement of the incident
photon energy deduced from the energy of the positron measured in the
small-angle positron calorimeter, to that calculated from energy
deposits in the uranium calorimeter.  The fivefold increase in
statistics in 1994 allowed the calorimeter energy scale to be studied
in more detail than before, and these studies revealed a $(6 \pm 3)\%$
difference between data and MC. This difference was removed in the
present analysis ( in the 1993 analysis~\cite{samerapdijet}, this
difference was not corrected, but the possibility of such a
discrepancy was allowed for in the systematic errors). Studies using
jet photoproduction events allow us to assign an uncertainty of 5\% to
the calorimeter response for the jets used in this
analysis~\cite{theses}.  Other studies have shown that of this 5\%
uncertainty, 3\% arises from the absolute energy
scale of the calorimeter~\cite{escale}.

After the jet energy correction, events are required to have at least
two jets with $\ETJ~\ge~6$~GeV, $-1.375 < \ETAJ <1.875$, and 
$|\DETA| < 0.5$. The MC gives a good description of the $|\DETA|$
distribution around this region.  For events with three or more jets,
the two highest $\ETJ$ jets are used to calculate all jet-related
event properties. This procedure is also employed later in all the
theoretical and MC predictions shown.

After these cuts about 25000 events remain, of which about 20\% have
$\xgo \ge 0.75$ (the exact number depending upon the jet
algorithm). Events with a third jet which passes the $\ETJ$ and
$\ETAJ$ cuts comprise about 8\% of the final sample. Of all events,
22\% have their scattered positrons detected in the small-angle
tagger, which is the fraction expected for a sample of photoproduction
events~\cite{ZEUSdir+inc}. No event from unpaired $e$ or $p$ bunches
survives the selection cuts, implying that the non-$ep$ background is
negligible.  The contamination from events with photon virtualities
greater than 4~GeV$^2$ is estimated using simulated DIS events.  This
contribution is much smaller than the statistical errors and is
therefore not subtracted from the data.

\section{Comparison with Monte Carlo Simulation}

\begin{figure}
\psfig{figure=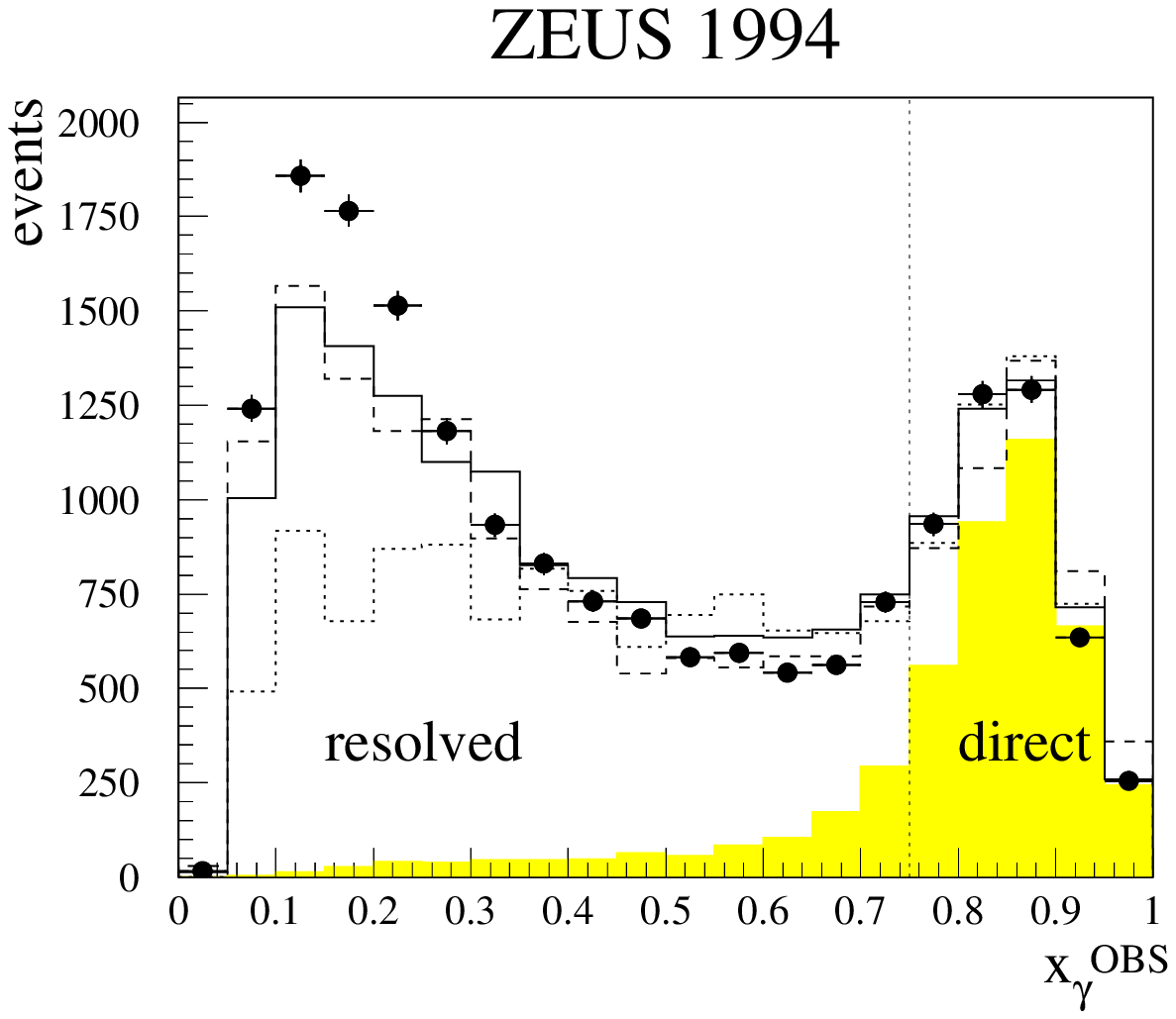}
\caption{ $\xgo$ distribution for KTCLUS jets with $\ETJ > 6$~GeV,
$-1.375 < \ETAJ <1.875$ and 
$|\DETA| < 0.5$,
where $\xgo$ is calculated using corrected variables.  The ZEUS 1994
data (black dots) are compared to the results of the HERWIG with MI
(solid line) and without (dotted line) and PYTHIA with MI (dashed
line) event generators after full detector simulation and scaling of
MC cross sections (see text). The HERWIG cross section for
resolved processes has been scaled by a factor of 1.2 with respect to
the direct.  The equivalent scaling for PYTHIA is the same.  Only
statistical errors are shown and in some cases are smaller than the
black dots. The shaded area represents the direct process HERWIG MC
events.
\label{f:fig1}}
\end{figure}

In Fig.~\ref{f:fig1} the $\xgo$ distribution of the ZEUS data selected
using the KTCLUS algorithm with (corrected) $\ETJ > 6$ GeV,
$-1.375 < \ETAJ <1.875$, and 
$|\DETA| < 0.5$ is shown
(black dots). This $\xgo$ is determined by using the corrected jet
energies and corrected $y_{JB}$.  The correction to $y_{JB}$ is
determined using MC generated events, by comparing
$y_{JB}$ to the true $y$, as a function of the $\xgo$ calculated using
uncorrected variables.  The peak at high $\xgo$ due to direct photon
processes and the rise at low $\xgo$ due to resolved photon processes
are both clearly visible. The sharp fall off for $\xgo < 0.1$ is a
result of the $\ETJ$ and $\ETAJ$ kinematic cuts.

The data are compared to the results of two LO QCD-based MC simulation
programs, HERWIG58~\cite{HERWIG} (solid line) and
PYTHIA57~\cite{PYTHIA} (dashed line).  All the MC events have been
passed through a detailed simulation of the ZEUS detector and through
the same jet energy correction procedure as was applied to the data.
The GRV~\cite{GRV} parton distributions are used for the photon and
the MRSA~\cite{MRSA} parton distributions are used for the proton. The
simulation programs are based on LO QCD calculations for the hard
scatter and include parton showering and hadronisation effects.  The
minimum transverse momentum of the partonic hard scatter ($\ptmin$) is
set to 2.5~GeV in both HERWIG and PYTHIA.  For both programs the
direct and resolved photon processes are generated separately. 

In the case of the resolved processes multiparton interactions (MI)
are included~\cite{MI,jimmy} as an attempt to simulate the energy 
from additional softer scatters (`underlying
event'), in both the dashed PYTHIA curve and the solid HERWIG curve.
This has been shown to improve the simulation of the energy flow
around the jet axis~\cite{MIexp}.

In order to obtain the best agreement with the data the normalisations
of the two processes were determined by allowing them to vary
independently and fitting to the uncorrected $\xgo$ distribution.  As
a result, the cross section from HERWIG for resolved processes was
scaled by 1.2 with respect to the direct. The ratio of direct and
resolved contributions using this scaled cross section was then 0.12,
to be compared with 0.15 when using the unscaled cross sections within
HERWIG.  For PYTHIA the equivalent scale factor for the resolved cross
section, and the ratio of direct and resolved, were found to be the
same as for HERWIG within the precision quoted here.

The dotted line shows the distribution for HERWIG without MI. For the
MI, models based upon the independent statistical replication of
scatters (eikonal models) are used which allow the generation of
additional independent partonic scatters (with transverse momenta
above \mbox{$\ptmin$ = 2.5 GeV} for HERWIG and 1.4 GeV for PYTHIA) in
resolved photon events. For HERWIG the average number of scatters for
events generated with these parameters is 1.05 and for PYTHIA it is
1.66. One effect of MI is to increase the number of events at low
$\xgo$.  However, even after the inclusion of MI with these
parameters, the data still lie above the simulation \mbox{at low $\xgo$.}

\begin{figure}
\psfig{figure=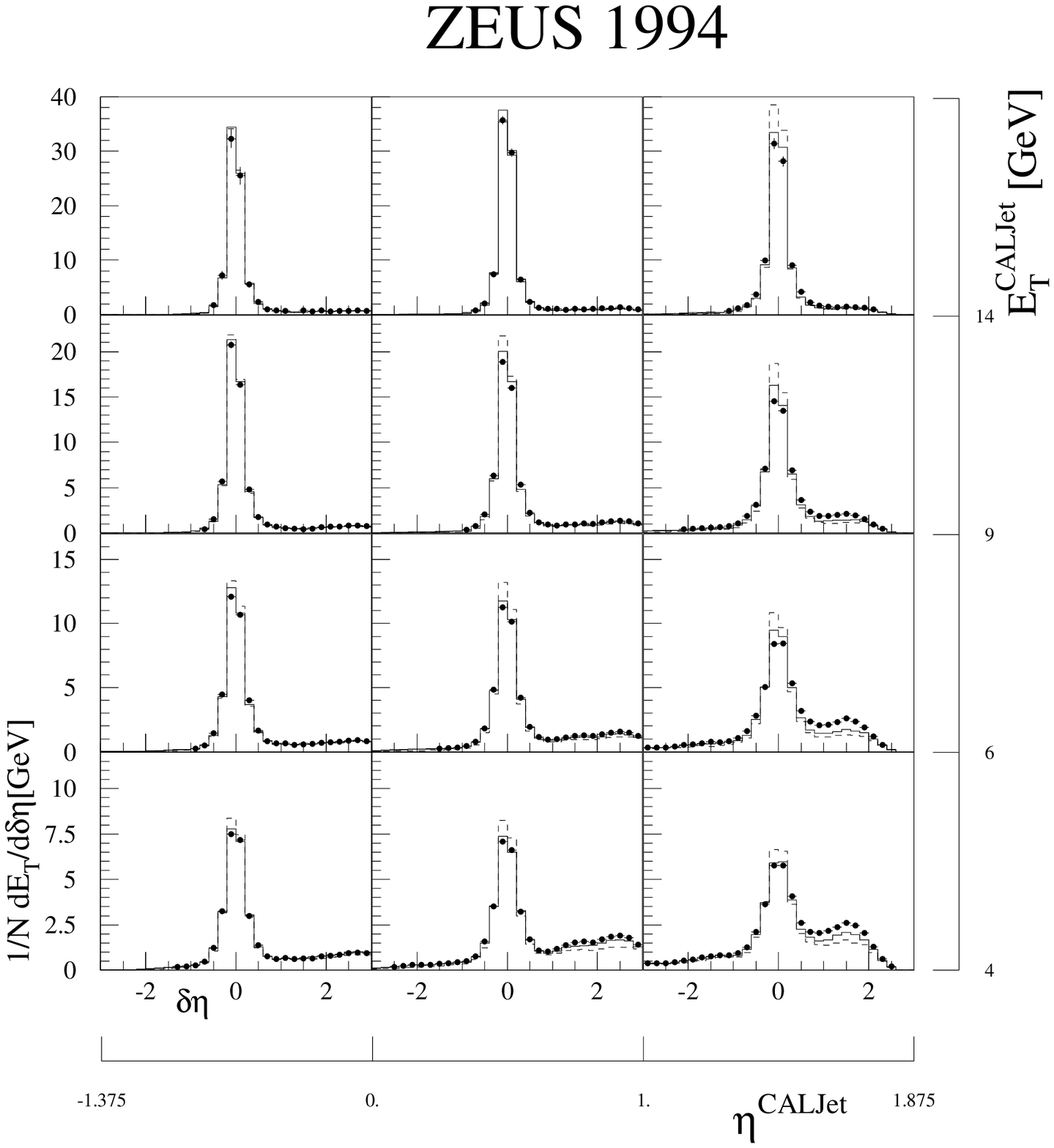,width={\textwidth}}
\caption{Uncorrected transverse energy flow $(1/N) dE_T/d\delta\eta$
around the jet axis, for cells within one radian in $\phi$ of the jet
axis, for KTCLUS 
jets binned in $\ETCJ$ and $\ETACJ$ (black dots), where
$\delta\eta=\eta^{CELL}-\eta^{CALjet}$.
The HERWIG MC with (solid line) and without (dashed line)
multiparton interactions are shown for comparison. 
Only statistical errors are shown.
\label{f:fig2}}
\end{figure}

The uncorrected transverse energy flow around the jets is shown in
Fig.~\ref{f:fig2}, for events in various bins of $\ETCJ$ and $\ETACJ$
for KTCLUS jets, and is compared to the distributions from the HERWIG
MC both with and without MI after full simulation of the detector.
The jet profiles are described reasonably well by the MC with MI for
most of the kinematic range, although there is still a tendency for MC
jets to have too much energy inside the central region and too little energy
outside this region, particularly for low $E_T$ jets
in the forward region. This tendency is significantly stronger for MC
samples which do not include multiparton interactions.  However, we do
not rule out the possibility that other models for the underlying event,
or different MI parameters not investigated here, may provide a
similar or better description of the data.

\section{Resolution and Systematics}

The resolution of the kinematic variables has been studied by
comparing, in the MC simulation, jets reconstructed from final state
particles (hadron jets) with jets reconstructed from the energies
measured in the calorimeter (detector jets), and by comparing
$y_{JB}$ with the true $y$.

The distribution of the difference in $\ETAB$ between the hadron and
detector jets has a mean of zero, a width of 0.15 units and depends
weakly on $\ETAB$, exhibiting shifts of about 0.05 units close to the
boundaries between the BCAL and the FCAL or RCAL. The resolution in
$\xgo$ is 8\% at $\xgo = 0.75$.  For $\ETJ$ and $y$, the resolutions
are 15\% and 0.09 units, respectively.

The jet cross sections presented in this analysis refer to jets in the
hadronic final state.  The MC samples have been used to correct the
data for the inefficiencies of the trigger and selection cuts and for
migrations caused by detector effects. The correction factors 
are calculated as the ratio
$N_{\rm true}/N_{\rm rec}$ in each bin.  $N_{\rm true}$ is the number
of events generated in the bin and $N_{\rm rec} $ is the number of
events reconstructed in the bin after detector smearing and all
experimental cuts. The final bin-by-bin
correction factors are between 0.5 and 1.5 for all the cross sections
measured. The dominant effect arises from migrations over the $\ETJ$
threshold. 

The sensitivity of the measured cross sections to the selection cuts
has been investigated by varying the cuts on the reconstructed
variables in the data and HERWIG MC samples and re-evaluating the
cross sections~\cite{theses}. In addition, the cross sections were
re-evaluated using a ratio of the direct and resolved contributions
derived from the cross sections from HERWIG without additional scaling
(direct/resolved=0.15), and by using the PYTHIA sample.  They were
also evaluated using the HERWIG model with and without multiparton
interactions.  These effects are included as systematic errors on the
cross sections, and are correlated to some extent.  The possibility
that the detector simulation may incorrectly simulate the detector
energy response by up to $\pm 5$\% has also been considered, as
mentioned in section~\ref{sec-ev-sel}. This effect is added in
quadrature to the overall normalisation error of $1.5\%$ arising from
the uncertainty in the measurement of the integrated luminosity. This
principal correlated uncertainty is indicated in the figures as a
shaded band and should be added to the other systematic errors to give
the overall uncertainty.

\section{Results}

The measured cross sections are now discussed and compared to
theoretical expectations. The cross section is first measured over
the whole $\xgo$ region and its shape is compared to that of MC
expectations.  This cross section includes $\xgo$ values down to 0.05,
the lowest value allowed by the other kinematic cuts.
At the lower
values of $\xgo$, the jet profiles and Fig.~\ref{f:fig1} indicate
discrepancies between the data and the MC simulations.  Nevertheless,
this cross section remains interesting as its shape is less biased by
kinematic cuts than those of the cross sections to be discussed in
section~7.2. We compare the shape to MC simulations which include
models for MI, parton showering and hadronisation, but have large
scale dependences due to the fact that they include only LO matrix
elements.

Next, $\xgo$ cuts are applied to select regions where contributions
arising from an underlying event - which may be responsible for the
low-$\xgo$ discrepancy in Fig.~\ref{f:fig1} - are reduced and hence
NLO QCD can be expected to provide a better description of the jet
production process.
The cross sections measured here in the hadronic final state are
compared to NLO QCD calculations of partonic cross sections.  These
calculations have a reduced scale dependence but do not include parton
showering beyond a single branching. MI and hadronization effects are
also not included since
no theoretical estimation of these two contributions is yet available
for these calculations. This uncertainty is not considered in
the following comparisons.

\subsection{Cross Sections without ${\boldmath \xgo}$ Cuts}

\begin{figure}
\epsfig{file=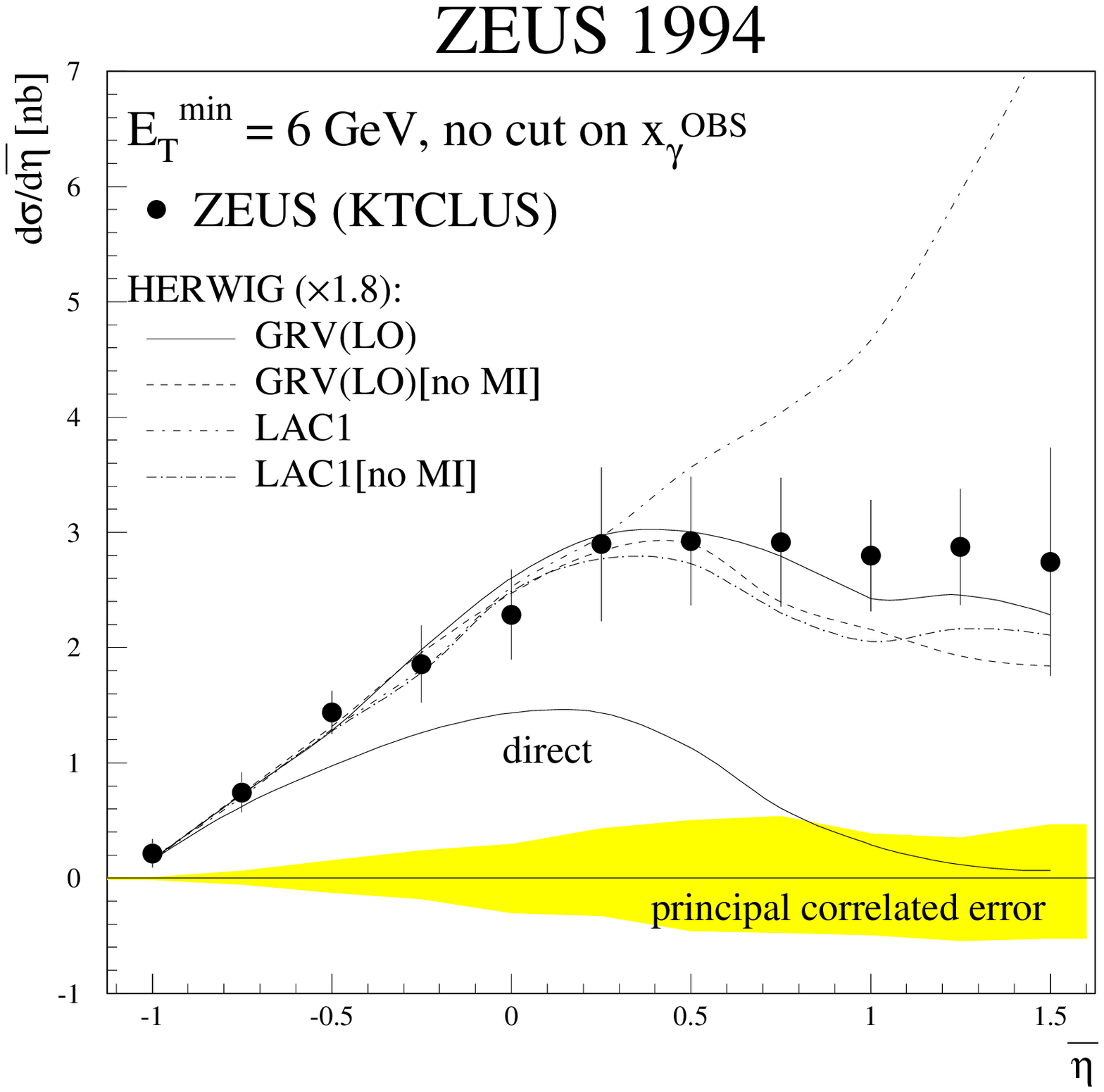,width={\textwidth}}
\caption{ $d\sigma/d\ETAB$ for $ep \rightarrow  e +$ dijets $+ X$ in the
range $|\DETA| < 0.5$, $0.2 < y < 0.8$ and for virtualities of the
exchanged photon $< 4$~GeV$^2$ and for $\ETJ$ integrated above $\ETJM
= 6$~GeV.  The cross section is measured using the KTCLUS algorithm
and is compared to the expectations of various HERWIG MC simulations
(see text).  The errors bars represent the combined systematic and
statistical uncertainty, excluding the principal correlated
uncertainties which are shown in the shaded band, see text.
\label{f:fig3}}
\end{figure}

The cross section $d\sigma/d\ETAB$ for $ep \rightarrow e + $ dijets $+
X$ in the range $|\DETA| < 0.5$, $0.2 < y < 0.8$ and for virtualities
of the exchanged photon less than $4$~GeV$^2$ is shown in
Fig.~\ref{f:fig3} and given in table~\ref{xs_kt_al_tab} for the KTCLUS
algorithm, requiring $\ETJ > 6$~GeV.

The cross section rises from around 0.2~nb per unit of pseudorapidity
at $\ETAB = -1$ to around 3~nb per unit of pseudorapidity for $\ETAB >
0.25$. The data may be compared with the predictions of the HERWIG MC
using the direct/resolved ratio of 0.15 given by HERWIG.  While the
simulation can describe the shape of the cross section, these
predictions fail to describe the overall normalisation, requiring an
ad hoc multiplicative scale factor of about 1.8 to agree with the
data. Such a factor is not unreasonable given the scale dependence of
the MC.  Fig.~\ref{f:fig3} shows various predictions of the HERWIG MC
after including the factor of 1.8.  With the value of $\ptmin =
2.5$~GeV used here, the data slightly favour the GRV parton
distribution~\cite{GRV} with MI. The LAC1~\cite{LAC1} or the GRV distribution
without MI also gives reasonable description of the data. However
the LAC1 distribution with MI is ruled out.

The effect of MI in the simulations is a strong function
of the choice of the photon parton distributions, in particular the
gluon component, which is where the major difference between LAC1 and
GRV lies. Additionally, it should be noted that the effect of MI is
also a strong function of the choice of $\ptmin$~\cite{jimmy}.  No
comparison is presented here with NLO perturbative QCD calculations
since they do not include MI. These comparisons are performed in the
next subsection, after applying $\xgo$ cuts to reduce such effects.

\subsection{Cross Sections with ${\boldmath \xgo}$ Cuts}

\begin{figure}
\begin{center}
\psfig{figure=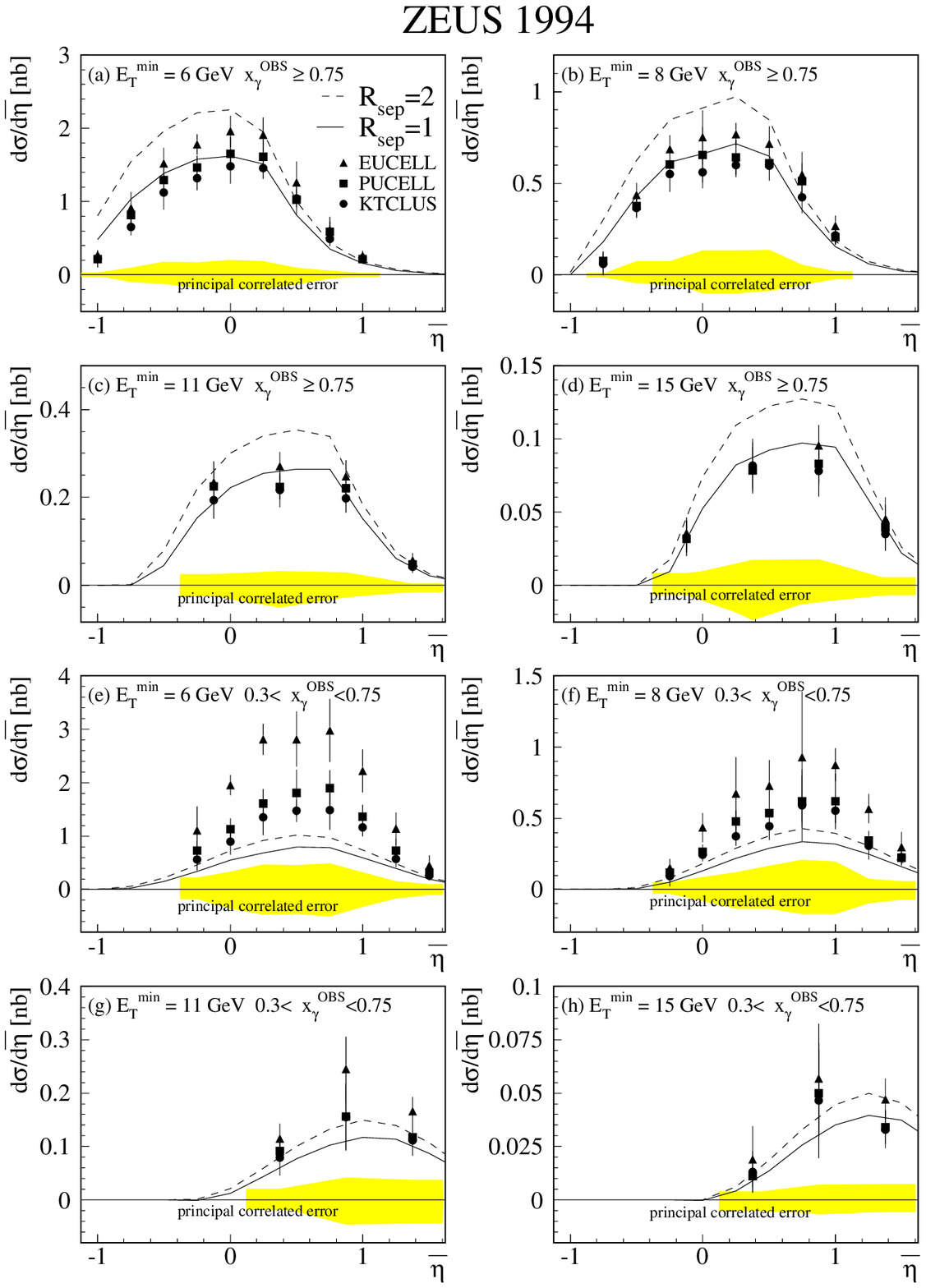}
\caption{ $d\sigma/d\ETAB$ for $ep \rightarrow e +$ dijets $+ X$ in
the range $|\DETA| < 0.5$, $0.2 < y < 0.8$ and for virtualities of the
exchanged photon $< 4$~GeV$^2$ and for $\ETJM$ = 6, 8, 11 and 15
GeV. Figures (a-d) are the cross sections measured in the range $\xgo
\ge 0.75$; figures (e-h) are for the range $0.3 < \xgo < 0.75$. The
cross sections are measured using three different jet algorithms and
are compared to NLO QCD calculations using $R_{sep}$ = 1 (solid
curves) and $R_{sep}$ = 2 (dashed curves), see text for details. The
errors bars represent the combined systematic and statistical
uncertainty, excluding the principal correlated uncertainties, which
are shown in the shaded band (see text). The band indicates the
maximum uncertainty for the three jet finders. The individual
uncertainty for each jet finder is given in the table.
\label{f:fig4}}
\end{center}
\end{figure}

Two regions have been selected:
\begin{enumerate}
\item $\xgo \ge 0.75$: direct photoproduction.
\item $0.3 < \xgo < 0.75$: resolved photoproduction excluding
the low-$\xgo$ region.
\end{enumerate}
For each $\xgo$ region, the cross sections $d\sigma/d\ETAB$ for $ep
\rightarrow e +$ dijets $+ X$ in the range $|\DETA| < 0.5$, $0.2 < y <
0.8$ and for virtualities of the exchanged photon $< 4$~GeV$^2$ are
measured for four different values of the $\ETJ$ threshold, $\ETJM$ =
6, 8, 11 and 15~GeV. 
The cross sections are measured for the three different jet algorithms
discussed in section~3. The results are given in Tables~\ref{xs_pu_di_tab}
to~\ref{xs_kt_re_tab} and are displayed in Fig.~\ref{f:fig4} together 
with the results of the NLO QCD calculation from Klasen and
Kramer~\cite{kk} using the NLO GRV~\cite{GRV} parton distributions for
the photon and the CTEQ3M~\cite{CTEQ} parton distributions for the
proton and employing two different values of the $\rsep$ parameter:
$\rsep = 1$ (solid curve) and $\rsep~=~2$ (dashed curve). Since the
jets may be accompanied by other soft gluons (outside the jets), there
is a potential problem when the two jets have the same $\ETJ$. The
infrared singularity associated with summing the soft gluon
contributions is usually cancelled by the singularity coming from the
one-loop contribution. For two jets with the same $\ETJ$, some of the
phase space for the soft gluon terms is restricted and an incomplete
cancellation may occur in some calculations. As a consequence, Klasen
and Kramer~\cite{kk} have allowed the second jet to have an $\ETJ$
less than $\ETJM$ if the third (unobserved gluon) jet has a transverse
energy of less than 1~GeV.  However, the cross section is then
sensitive to changes in the value used for this cut on the third jet.
Harris and Owens~\cite{owens} have applied a low cutoff on the energy
of the very soft gluons and found that the dependence of the cross
sections on the value of the low energy cutoff used is much less than
the quoted errors on the data. These different approaches account for
the differences between the theory curves shown later.

The photoproduction cross section for $\xgo \ge 0.75$ and $\ETJM = 6$~GeV
(Fig.~\ref{f:fig4}a) rises from around 0.2~nb at $\ETAB = -1$ to a
maximum value of around 1.8~nb at $\ETAB = 0$, decreasing back to
0.2~nb by $\ETAB = 1$. This decrease arises from the cutoff on the
minimum $\ETJ$ and the cuts on $y$. The EUCELL jet cross sections are
systematically higher than the PUCELL cross sections, which in turn
are slightly above the KTCLUS cross sections. This behaviour is
qualitatively similar for the higher $\ETJM$ cross sections
(Figs.~\ref{f:fig4}b-d), where the maximum value of the cross section
falls and occurs at steadily higher $\ETAB$ as the minimum $\ETJ$ cut
increases. The PUCELL and KTCLUS cross sections are in good agreement
with the NLO curve calculated with $\rsep = 1$ for all $\ETAB$ and for
all four $\ETJ$ thresholds, except for the most negative values of
$\ETAB$ in the lower $\ETJM$ cross sections, where the trend is for
the calculation to lie above the data. The $\rsep = 2$ curve lies
above all the data at most values of $\ETAB$. In the data the
separation between EUCELL, PUCELL and KTCLUS becomes less significant
at higher $\ETJ$. However, the separation between the two theory
curves remains significant.

The photoproduction cross section for $0.3 < \xgo < 0.75$ and $\ETJM = 6$~GeV
(Fig.~\ref{f:fig4}e) rises from around 0.8~nb at $\ETAB = -0.25$ to a
maximum value of 1.5~nb for PUCELL and KTCLUS, and of 3~nb for EUCELL,
at $\ETAB = 0$, followed by a decrease back to 0.2~nb by $\ETAB =
1.5$. The EUCELL jet cross sections are again systematically higher
than the PUCELL cross sections which are again slightly above those
for KTCLUS. This behaviour is once more qualitatively similar for the
higher $\ETJ$ cross sections (Figs.~\ref{f:fig4}f-h), where the
maximum value of the cross section falls and occurs at steadily higher
$\ETAB$. In the data the separation between EUCELL and the two other
jet algorithms is larger than in the direct case - a factor of two at
the lowest $\ETJ$ values - but again becomes less significant at
higher $\ETJ$. In the theory, the differences between the curves with
different $\rsep$ again show the same trend as the data, with the
$\rsep = 2$ curves being higher than those for $\rsep = 1$.  However,
for the cross sections with $\ETJM = 6$~GeV and 8~GeV, the NLO QCD
curves lie significantly below the data. For higher $\ETJM$ values the
calculations are broadly consistent with the data.  

\begin{figure}
\centering
\psfig{figure=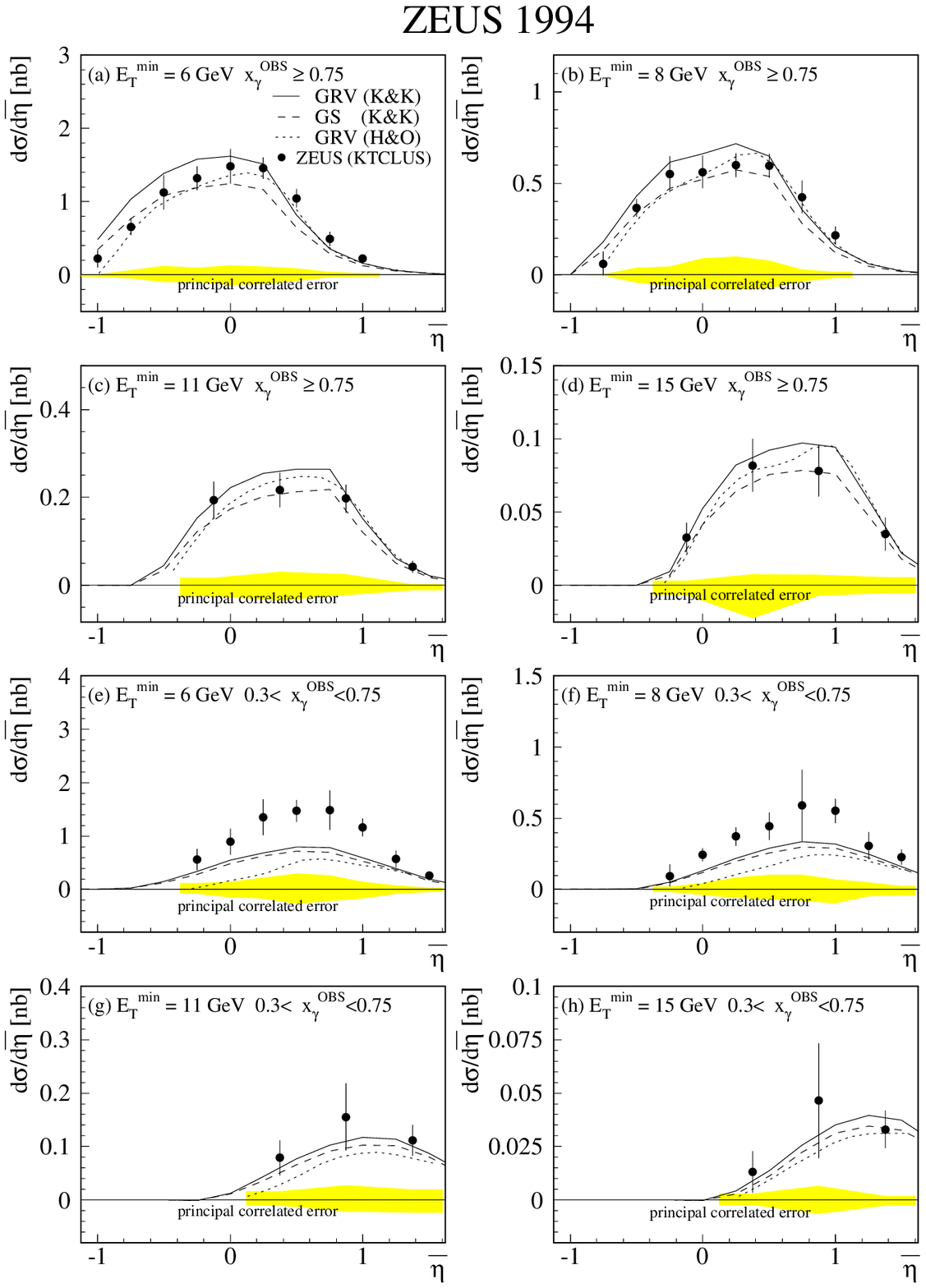}
\caption{ $d\sigma/d\ETAB$ for $ep \rightarrow e +$ dijets $+ X$ in
the range $|\DETA| < 0.5$, $0.2 < y < 0.8$ and for virtualities of the
exchanged photon $< 4$~GeV$^2$ and for $\ETJM$ = 6, 8, 11 and 15~GeV.
Figures (a-d) are the cross sections measured in the range $\xgo \ge
0.75$; figures (e-h) are for the range $0.3 < \xgo < 0.75$. The cross
sections are measured using the KTCLUS jet algorithm and are compared
to NLO QCD calculations from Klasen and Kramer, for two different
parton distributions in the photon , GRV and GS, and from Harris 
and Owens using GRV
(see text for details).  All three calculations use $R_{sep}$ = 1.
The errors bars represent the combined systematic and statistical
uncertainty, excluding the principal correlated uncertainties which
are shown in the shaded band (see text).
\label{f:fig5}}
\end{figure}

In Fig.~\ref{f:fig5} the KTCLUS jet cross sections are shown again,
with Klasen and Kramer's NLO QCD calculations (with $\rsep = 1$)
employing two different parton distribution functions for the photon -
namely those of NLO GRV(solid curves~\cite{GRV}) and GS (dashed
curves~\cite{GS}). It can be seen that the agreement is in general
good for both distribution functions, except in the two lowest $\ETJ$
regions of the resolved cross section (Fig~\ref{f:fig5}e,f) where the
QCD calculations are significantly below the data.  Perhaps
surprisingly, the difference between the photon parton distributions
is largest in the direct photoproduction region.  This is due to
differences between the quark distributions in the photon for
$x_\gamma > 0.8$, where they are poorly constrained by photon
structure function measurements at $e^+e^-$ colliders.  These
differences persist at high $\ETJ$. Also shown in Fig.~\ref{f:fig5} is
a NLO QCD calculation (again with $\rsep = 1$) from Harris and Owens
using NLO GRV $\overline{\rm MS}$ for the photon, and NLO CTEQ4
$\overline{\rm MS}$ for the proton. At high $\xgo$ there is again good
agreement with the measurements, but at low $\xgo$ the disagreement in
the two lowest $\ETJ$ regions is large.  At higher $\ETJ$ values, the
data and calculations are consistent.

\section{Conclusions} 

Photoproduced dijet cross sections $d\sigma/d\ETAB$ have been measured
in the hadronic final state for different kinematic regions
and are found to be consistent with the general expectations of QCD,
in the sense that both resolved and direct processes are observed in
the data.

Quantitatively, it is found that Monte Carlo simulations both with and
without multiparton interactions are capable of describing the $\ETAB$
dependence of the cross section when no $\xgo$ cuts are applied,
although simulations which use multiparton interactions to simulate an
underlying event are slightly favoured and also give a better
description of the jet profiles.

The measured cross sections vary by up to a factor of two when
different cone or clustering algorithms are used for the definition of
jets.  This behaviour is similar to that predicted 
from the theoretical calculations by choosing the
$\rsep$ parameter in order to reproduce the 
different jet algorithms.

Comparison of the direct photon cross sections ($\xgo > 0.75$) with
NLO QCD calculations shows good agreement in both shape and magnitude
over a wide range of $\ETJ$ and $\ETAJ$ and for the three different
jet definitions.  It also displays a sensitivity to the photon
structure at \mbox{large $x_{\gamma}$.}

Calculations for the resolved photon cross sections in the region $0.3
< \xgo < 0.75$ which include jets with 6~GeV$ < \ETJ < 11$~GeV are
found to lie below the data. However, for higher jet energies the
calculations are consistent with the data.

\begin{table}
\begin{center}
\begin{tabular}{|r|c|c|c|cc|r@{\hspace{3ex}}|}
\hline
\multicolumn{1}{|c|}{$\bar{\eta}$} & \rule[-3ex]{0ex}{6ex}$\displaystyle\frac{d\sigma}{d\bar{\eta}}$~[nb] & stat.~[nb] & syst.~[nb] & \multicolumn{2}{|c|}{corrl. syst.~[nb]}   \\
\hline
\hline
\multicolumn{6}{|l|}{\rule[-2ex]{0ex}{5ex}$E_T^{min}>6~{\rm GeV}$}\\
\hline
$-1.000$ &   0.22 &   0.02 &   0.12 & $+0.01$ & $-0.01$ \\
$-0.750$ &   0.75 &   0.05 &   0.16 & $+0.06$ & $-0.05$ \\
$-0.500$ &   1.44 &   0.08 &   0.17 & $+0.15$ & $-0.13$ \\
$-0.250$ &   1.86 &   0.08 &   0.32 & $+0.24$ & $-0.18$ \\
$ 0.000$ &   2.29 &   0.09 &   0.38 & $+0.30$ & $-0.30$ \\
$ 0.250$ &   2.90 &   0.10 &   0.66 & $+0.43$ & $-0.33$ \\
$ 0.500$ &   2.92 &   0.10 &   0.55 & $+0.50$ & $-0.46$ \\
$ 0.750$ &   2.91 &   0.11 &   0.55 & $+0.54$ & $-0.47$ \\
$ 1.000$ &   2.80 &   0.11 &   0.47 & $+0.38$ & $-0.49$ \\
$ 1.250$ &   2.87 &   0.12 &   0.49 & $+0.35$ & $-0.54$ \\
$ 1.500$ &   2.74 &   0.11 &   0.98 & $+0.46$ & $-0.52$ \\
\hline
\hline
\multicolumn{6}{|l|}{\rule[-2ex]{0ex}{5ex}$E_T^{min}>8~{\rm GeV}$}\\
\hline
$-0.750$ &   0.06 &   0.01 &   0.06 & $+0.01$ & $-0.01$ \\
$-0.500$ &   0.38 &   0.04 &   0.02 & $+0.04$ & $-0.03$ \\
$-0.250$ &   0.65 &   0.05 &   0.08 & $+0.07$ & $-0.08$ \\
$ 0.000$ &   0.81 &   0.05 &   0.09 & $+0.13$ & $-0.11$ \\
$ 0.250$ &   0.97 &   0.06 &   0.11 & $+0.19$ & $-0.13$ \\
$ 0.500$ &   1.07 &   0.06 &   0.17 & $+0.18$ & $-0.15$ \\
$ 0.750$ &   1.16 &   0.06 &   0.20 & $+0.17$ & $-0.14$ \\
$ 1.000$ &   0.99 &   0.06 &   0.12 & $+0.15$ & $-0.17$ \\
$ 1.250$ &   0.82 &   0.06 &   0.12 & $+0.12$ & $-0.18$ \\
$ 1.500$ &   0.71 &   0.05 &   0.08 & $+0.13$ & $-0.15$ \\
\hline
\hline
\multicolumn{6}{|l|}{\rule[-2ex]{0ex}{5ex}$E_T^{min}>11~{\rm GeV}$}\\
\hline
$-0.125$ &   0.21 &   0.02 &   0.02 & $+0.02$ & $-0.02$ \\
$ 0.375$ &   0.29 &   0.02 &   0.02 & $+0.05$ & $-0.04$ \\
$ 0.875$ &   0.37 &   0.02 &   0.06 & $+0.06$ & $-0.05$ \\
$ 1.375$ &   0.23 &   0.02 &   0.02 & $+0.03$ & $-0.05$ \\
\hline
\hline
\multicolumn{6}{|l|}{\rule[-2ex]{0ex}{5ex}$E_T^{min}>15~{\rm GeV}$}\\
\hline
$-0.125$ &  0.033 &  0.008 &  0.005 & $+0.003$ & $-0.004$ \\
$ 0.375$ &  0.093 &  0.012 &  0.036 & $+0.011$ & $-0.023$ \\
$ 0.875$ &  0.126 &  0.014 &  0.027 & $+0.014$ & $-0.014$ \\
$ 1.375$ &  0.079 &  0.010 &  0.016 & $+0.012$ & $-0.010$ \\
\hline
\end{tabular}
\caption{\label{xs_kt_al_tab}The cross sections for KTCLUS for the
  whole $\xgo$ range.  The third and fourth columns represent the 
  statistical and systematic uncertainties, excluding the principal
  correlated uncertainties which are shown in the fifth column, see
  text.}
\end{center}
\end{table}
\begin{table}
\begin{center}
\begin{tabular}{|r|c|c|c|cc|}
\hline
\multicolumn{1}{|c|}{$\bar{\eta}$} & \rule[-3ex]{0ex}{6ex}$\displaystyle\frac{d\sigma}{d\bar{\eta}}$~[nb] & stat.~[nb] & syst.~[nb] & \multicolumn{2}{|c|}{corrl. syst.~[nb]} \\
\hline
\hline
\multicolumn{6}{|l|}{\rule[-2ex]{0ex}{5ex}$E_T^{min}>6~{\rm GeV}$}\\
\hline
$-1.000$ &   0.22 &   0.03 &   0.05 & $+0.02$ & $-0.02$ \\
$-0.750$ &   0.81 &   0.06 &   0.09 & $+0.04$ & $-0.08$ \\
$-0.500$ &   1.29 &   0.08 &   0.15 & $+0.14$ & $-0.09$ \\
$-0.250$ &   1.47 &   0.08 &   0.14 & $+0.13$ & $-0.15$ \\
$ 0.000$ &   1.65 &   0.08 &   0.21 & $+0.14$ & $-0.14$ \\
$ 0.250$ &   1.61 &   0.08 &   0.20 & $+0.16$ & $-0.15$ \\
$ 0.500$ &   1.03 &   0.06 &   0.20 & $+0.07$ & $-0.09$ \\
$ 0.750$ &   0.59 &   0.04 &   0.13 & $+0.05$ & $-0.04$ \\
$ 1.000$ &   0.21 &   0.02 &   0.03 & $+0.02$ & $-0.02$ \\
\hline
\hline
\multicolumn{6}{|l|}{\rule[-2ex]{0ex}{5ex}$E_T^{min}>8~{\rm GeV}$}\\
\hline
$-0.750$ &   0.08 &   0.01 &   0.03 & $+0.01$ & $-0.01$ \\
$-0.500$ &   0.38 &   0.04 &   0.05 & $+0.02$ & $-0.04$ \\
$-0.250$ &   0.60 &   0.05 &   0.05 & $+0.04$ & $-0.05$ \\
$ 0.000$ &   0.66 &   0.05 &   0.14 & $+0.10$ & $-0.07$ \\
$ 0.250$ &   0.64 &   0.05 &   0.08 & $+0.11$ & $-0.09$ \\
$ 0.500$ &   0.61 &   0.05 &   0.08 & $+0.08$ & $-0.08$ \\
$ 0.750$ &   0.51 &   0.04 &   0.09 & $+0.03$ & $-0.04$ \\
$ 1.000$ &   0.21 &   0.02 &   0.03 & $+0.01$ & $-0.02$ \\
\hline
\hline
\multicolumn{6}{|l|}{\rule[-2ex]{0ex}{5ex}$E_T^{min}>11~{\rm GeV}$}\\
\hline
$-0.125$ &   0.23 &   0.02 &   0.04 & $+0.01$ & $-0.02$ \\
$ 0.375$ &   0.22 &   0.02 &   0.02 & $+0.03$ & $-0.03$ \\
$ 0.875$ &   0.22 &   0.02 &   0.03 & $+0.02$ & $-0.03$ \\
$ 1.375$ &   0.05 &   0.01 &   0.02 & $+0.01$ & $-0.01$ \\
\hline
\hline
\multicolumn{6}{|l|}{\rule[-2ex]{0ex}{5ex}$E_T^{min}>15~{\rm GeV}$}\\
\hline
$-0.125$ &  0.032 &  0.008 &  0.009 & $+0.008$ & $-0.003$ \\
$ 0.375$ &  0.078 &  0.011 &  0.011 & $+0.012$ & $-0.024$ \\
$ 0.875$ &  0.083 &  0.012 &  0.006 & $+0.013$ & $-0.009$ \\
$ 1.375$ &  0.039 &  0.008 &  0.014 & $+0.004$ & $-0.007$ \\
\hline
\end{tabular}
\caption{\label{xs_pu_di_tab}The cross sections for PUCELL and
  $\xgo \ge 0.75$. The third and fourth columns represent the 
  statistical and systematic uncertainties, excluding the principal
  correlated uncertainties which are shown in the fifth column, see
  text.}
\end{center}
\end{table}
\begin{table}
\begin{center}
\begin{tabular}{|r|c|c|c|cc|}
\hline
\multicolumn{1}{|c|}{$\bar{\eta}$} & \rule[-3ex]{0ex}{6ex}$\displaystyle\frac{d\sigma}{d\bar{\eta}}$~[nb] & stat.~[nb] & syst.~[nb] & \multicolumn{2}{|c|}{corrl. syst.~[nb]} \\
\hline
\hline
\multicolumn{6}{|l|}{\rule[-2ex]{0ex}{5ex}$E_T^{min}>6~{\rm GeV}$}\\
\hline
$-0.250$ &   0.73 &   0.06 &   0.17 & $+0.15$ & $-0.09$ \\
$ 0.000$ &   1.13 &   0.07 &   0.19 & $+0.19$ & $-0.17$ \\
$ 0.250$ &   1.61 &   0.08 &   0.26 & $+0.26$ & $-0.25$ \\
$ 0.500$ &   1.81 &   0.09 &   0.43 & $+0.29$ & $-0.29$ \\
$ 0.750$ &   1.90 &   0.10 &   0.32 & $+0.31$ & $-0.34$ \\
$ 1.000$ &   1.37 &   0.09 &   0.20 & $+0.23$ & $-0.19$ \\
$ 1.250$ &   0.73 &   0.06 &   0.21 & $+0.10$ & $-0.10$ \\
$ 1.500$ &   0.32 &   0.03 &   0.06 & $+0.06$ & $-0.06$ \\
\hline
\hline
\multicolumn{6}{|l|}{\rule[-2ex]{0ex}{5ex}$E_T^{min}>8~{\rm GeV}$}\\
\hline
$-0.250$ &   0.12 &   0.02 &   0.04 & $+0.02$ & $-0.02$ \\
$ 0.000$ &   0.26 &   0.03 &   0.04 & $+0.05$ & $-0.03$ \\
$ 0.250$ &   0.48 &   0.04 &   0.06 & $+0.08$ & $-0.10$ \\
$ 0.500$ &   0.54 &   0.04 &   0.10 & $+0.11$ & $-0.09$ \\
$ 0.750$ &   0.62 &   0.05 &   0.18 & $+0.11$ & $-0.10$ \\
$ 1.000$ &   0.62 &   0.05 &   0.19 & $+0.12$ & $-0.12$ \\
$ 1.250$ &   0.35 &   0.04 &   0.05 & $+0.06$ & $-0.05$ \\
$ 1.500$ &   0.22 &   0.03 &   0.04 & $+0.04$ & $-0.04$ \\
\hline
\hline
\multicolumn{6}{|l|}{\rule[-2ex]{0ex}{5ex}$E_T^{min}>11~{\rm GeV}$}\\
\hline
$ 0.375$ &   0.09 &   0.01 &   0.02 & $+0.01$ & $-0.02$ \\
$ 0.875$ &   0.16 &   0.02 &   0.06 & $+0.03$ & $-0.03$ \\
$ 1.375$ &   0.12 &   0.01 &   0.02 & $+0.02$ & $-0.02$ \\
\hline
\hline
\multicolumn{6}{|l|}{\rule[-2ex]{0ex}{5ex}$E_T^{min}>15~{\rm GeV}$}\\
\hline
$ 0.375$ &  0.011 &  0.003 &  0.004 & $+0.004$ & $-0.002$ \\
$ 0.875$ &  0.050 &  0.009 &  0.018 & $+0.002$ & $-0.007$ \\
$ 1.375$ &  0.034 &  0.006 &  0.005 & $+0.003$ & $-0.005$ \\
\hline
\end{tabular}
\caption{\label{xs_pu_re_tab}The cross sections for PUCELL and
  $0.30<\xgo<0.75$. The third and fourth columns represent the 
  statistical and systematic uncertainties, excluding the principal
  correlated uncertainties which are shown in the fifth column, see
  text.}
\end{center}
\end{table}
\begin{table}
\begin{center}
\begin{tabular}{|r|c|c|c|cc|r@{\hspace{3ex}}|}
\hline
\multicolumn{1}{|c|}{$\bar{\eta}$} & \rule[-3ex]{0ex}{6ex}$\displaystyle\frac{d\sigma}{d\bar{\eta}}$~[nb] & stat.~[nb] & syst.~[nb] & \multicolumn{2}{|c|}{corrl. syst.~[nb]}   \\
\hline
\hline
\multicolumn{6}{|l|}{\rule[-2ex]{0ex}{5ex}$E_T^{min}>6~{\rm GeV}$}\\
\hline
$-1.000$ &   0.27 &   0.03 &   0.04 & $+0.02$ & $-0.01$ \\
$-0.750$ &   0.92 &   0.06 &   0.21 & $+0.09$ & $-0.10$ \\
$-0.500$ &   1.52 &   0.08 &   0.20 & $+0.18$ & $-0.12$ \\
$-0.250$ &   1.78 &   0.09 &   0.11 & $+0.17$ & $-0.17$ \\
$ 0.000$ &   1.96 &   0.09 &   0.20 & $+0.20$ & $-0.19$ \\
$ 0.250$ &   1.91 &   0.09 &   0.22 & $+0.18$ & $-0.19$ \\
$ 0.500$ &   1.27 &   0.07 &   0.27 & $+0.09$ & $-0.13$ \\
$ 0.750$ &   0.62 &   0.04 &   0.16 & $+0.06$ & $-0.05$ \\
$ 1.000$ &   0.27 &   0.03 &   0.05 & $+0.02$ & $-0.02$ \\
\hline
\hline
\multicolumn{6}{|l|}{\rule[-2ex]{0ex}{5ex}$E_T^{min}>8~{\rm GeV}$}\\
\hline
$-0.750$ &   0.08 &   0.01 &   0.05 & $+0.01$ & $-0.01$ \\
$-0.500$ &   0.44 &   0.04 &   0.05 & $+0.07$ & $-0.04$ \\
$-0.250$ &   0.69 &   0.05 &   0.05 & $+0.07$ & $-0.03$ \\
$ 0.000$ &   0.75 &   0.05 &   0.14 & $+0.13$ & $-0.10$ \\
$ 0.250$ &   0.77 &   0.05 &   0.03 & $+0.13$ & $-0.10$ \\
$ 0.500$ &   0.72 &   0.05 &   0.08 & $+0.13$ & $-0.09$ \\
$ 0.750$ &   0.55 &   0.04 &   0.12 & $+0.05$ & $-0.05$ \\
$ 1.000$ &   0.27 &   0.03 &   0.05 & $+0.02$ & $-0.02$ \\
\hline
\hline
\multicolumn{6}{|l|}{\rule[-2ex]{0ex}{5ex}$E_T^{min}>11~{\rm GeV}$}\\
\hline
$-0.125$ &   0.23 &   0.02 &   0.04 & $+0.02$ & $-0.02$ \\
$ 0.375$ &   0.27 &   0.02 &   0.02 & $+0.03$ & $-0.05$ \\
$ 0.875$ &   0.25 &   0.02 &   0.03 & $+0.03$ & $-0.03$ \\
$ 1.375$ &   0.06 &   0.01 &   0.01 & $+0.01$ & $-0.01$ \\
\hline
\hline
\multicolumn{6}{|l|}{\rule[-2ex]{0ex}{5ex}$E_T^{min}>15~{\rm GeV}$}\\
\hline
$-0.125$ &  0.036 &  0.009 &  0.007 & $+0.007$ & $-0.004$ \\
$ 0.375$ &  0.082 &  0.012 &  0.010 & $+0.017$ & $-0.017$ \\
$ 0.875$ &  0.095 &  0.013 &  0.006 & $+0.017$ & $-0.011$ \\
$ 1.375$ &  0.045 &  0.007 &  0.013 & $+0.004$ & $-0.006$ \\
\hline
\end{tabular}
\caption{\label{xs_eu_di_tab}The cross sections for EUCELL and
  $\xgo \ge 0.75$.  The third and fourth columns represent the 
  statistical and systematic uncertainties, excluding the principal
  correlated uncertainties which are shown in the fifth column, see
  text.}
\end{center}
\end{table}
\begin{table}
\begin{center}
\begin{tabular}{|r|c|c|c|cc|r@{\hspace{3ex}}|}
\hline
\multicolumn{1}{|c|}{$\bar{\eta}$} & \rule[-3ex]{0ex}{6ex}$\displaystyle\frac{d\sigma}{d\bar{\eta}}$~[nb] & stat.~[nb] & syst.~[nb] & \multicolumn{2}{|c|}{corrl. syst.~[nb]}   \\
\hline
\hline
\multicolumn{6}{|l|}{\rule[-2ex]{0ex}{5ex}$E_T^{min}>6~{\rm GeV}$}\\
\hline
$-0.250$ &   1.11 &   0.07 &   0.43 & $+0.21$ & $-0.18$ \\
$ 0.000$ &   1.95 &   0.10 &   0.16 & $+0.33$ & $-0.36$ \\
$ 0.250$ &   2.81 &   0.12 &   0.26 & $+0.47$ & $-0.47$ \\
$ 0.500$ &   2.82 &   0.12 &   0.51 & $+0.45$ & $-0.45$ \\
$ 0.750$ &   2.98 &   0.13 &   0.57 & $+0.49$ & $-0.49$ \\
$ 1.000$ &   2.22 &   0.12 &   0.38 & $+0.31$ & $-0.32$ \\
$ 1.250$ &   1.14 &   0.08 &   0.30 & $+0.15$ & $-0.16$ \\
$ 1.500$ &   0.45 &   0.04 &   0.19 & $+0.10$ & $-0.10$ \\
\hline
\hline
\multicolumn{6}{|l|}{\rule[-2ex]{0ex}{5ex}$E_T^{min}>8~{\rm GeV}$}\\
\hline
$-0.250$ &   0.15 &   0.02 &   0.06 & $+0.05$ & $-0.03$ \\
$ 0.000$ &   0.44 &   0.04 &   0.09 & $+0.09$ & $-0.08$ \\
$ 0.250$ &   0.68 &   0.05 &   0.25 & $+0.12$ & $-0.13$ \\
$ 0.500$ &   0.73 &   0.05 &   0.17 & $+0.17$ & $-0.13$ \\
$ 0.750$ &   0.93 &   0.06 &   0.45 & $+0.21$ & $-0.17$ \\
$ 1.000$ &   0.88 &   0.07 &   0.09 & $+0.19$ & $-0.17$ \\
$ 1.250$ &   0.57 &   0.05 &   0.09 & $+0.07$ & $-0.09$ \\
$ 1.500$ &   0.30 &   0.03 &   0.10 & $+0.06$ & $-0.07$ \\
\hline
\hline
\multicolumn{6}{|l|}{\rule[-2ex]{0ex}{5ex}$E_T^{min}>11~{\rm GeV}$}\\
\hline
$ 0.375$ &   0.11 &   0.01 &   0.03 & $+0.02$ & $-0.01$ \\
$ 0.875$ &   0.24 &   0.02 &   0.06 & $+0.04$ & $-0.05$ \\
$ 1.375$ &   0.17 &   0.02 &   0.02 & $+0.04$ & $-0.04$ \\
\hline
\hline
\multicolumn{6}{|l|}{\rule[-2ex]{0ex}{5ex}$E_T^{min}>15~{\rm GeV}$}\\
\hline
$ 0.375$ &  0.019 &  0.005 &  0.015 & $+0.004$ & $-0.005$ \\
$ 0.875$ &  0.057 &  0.009 &  0.024 & $+0.007$ & $-0.006$ \\
$ 1.375$ &  0.047 &  0.008 &  0.006 & $+0.007$ & $-0.005$ \\
\hline
\end{tabular}
\caption{\label{xs_eu_re_tab}The cross sections for EUCELL and
  $0.30<\xgo<0.75$. The third and fourth columns represent the 
  statistical and systematic uncertainties, excluding the principal
  correlated uncertainties which are shown in the fifth column, see
  text.}
\end{center}
\end{table}
\begin{table}
\begin{center}
\begin{tabular}{|r|c|c|c|cc|r@{\hspace{3ex}}|}
\hline
\multicolumn{1}{|c|}{$\bar{\eta}$} & \rule[-3ex]{0ex}{6ex}$\displaystyle\frac{d\sigma}{d\bar{\eta}}$~[nb] & stat.~[nb] & syst.~[nb] & \multicolumn{2}{|c|}{corrl. syst.~[nb]}   \\
\hline
\hline
\multicolumn{6}{|l|}{\rule[-2ex]{0ex}{5ex}$E_T^{min}>6~{\rm GeV}$}\\
\hline
$-1.000$ &   0.22 &   0.03 &   0.12 & $+0.01$ & $-0.02$ \\
$-0.750$ &   0.66 &   0.05 &   0.10 & $+0.06$ & $-0.05$ \\
$-0.500$ &   1.12 &   0.07 &   0.22 & $+0.12$ & $-0.10$ \\
$-0.250$ &   1.32 &   0.07 &   0.15 & $+0.10$ & $-0.09$ \\
$ 0.000$ &   1.48 &   0.07 &   0.22 & $+0.13$ & $-0.14$ \\
$ 0.250$ &   1.46 &   0.07 &   0.13 & $+0.11$ & $-0.12$ \\
$ 0.500$ &   1.05 &   0.06 &   0.11 & $+0.08$ & $-0.08$ \\
$ 0.750$ &   0.49 &   0.04 &   0.09 & $+0.04$ & $-0.03$ \\
$ 1.000$ &   0.22 &   0.03 &   0.05 & $+0.02$ & $-0.02$ \\
\hline
\hline
\multicolumn{6}{|l|}{\rule[-2ex]{0ex}{5ex}$E_T^{min}>8~{\rm GeV}$}\\
\hline
$-0.750$ &   0.06 &   0.01 &   0.06 & $+0.01$ & $-0.01$ \\
$-0.500$ &   0.36 &   0.04 &   0.03 & $+0.04$ & $-0.04$ \\
$-0.250$ &   0.55 &   0.05 &   0.09 & $+0.04$ & $-0.06$ \\
$ 0.000$ &   0.56 &   0.04 &   0.08 & $+0.09$ & $-0.07$ \\
$ 0.250$ &   0.60 &   0.04 &   0.05 & $+0.10$ & $-0.08$ \\
$ 0.500$ &   0.60 &   0.04 &   0.05 & $+0.08$ & $-0.07$ \\
$ 0.750$ &   0.42 &   0.04 &   0.08 & $+0.03$ & $-0.03$ \\
$ 1.000$ &   0.22 &   0.03 &   0.04 & $+0.01$ & $-0.02$ \\
\hline
\hline
\multicolumn{6}{|l|}{\rule[-2ex]{0ex}{5ex}$E_T^{min}>11~{\rm GeV}$}\\
\hline
$-0.125$ &   0.19 &   0.02 &   0.04 & $+0.02$ & $-0.02$ \\
$ 0.375$ &   0.22 &   0.02 &   0.03 & $+0.03$ & $-0.03$ \\
$ 0.875$ &   0.20 &   0.02 &   0.03 & $+0.02$ & $-0.02$ \\
$ 1.375$ &   0.04 &   0.01 &   0.01 & $+0.01$ & $-0.01$ \\
\hline
\hline
\multicolumn{6}{|l|}{\rule[-2ex]{0ex}{5ex}$E_T^{min}>15~{\rm GeV}$}\\
\hline
$-0.125$ &  0.033 &  0.008 &  0.006 & $+0.003$ & $-0.007$ \\
$ 0.375$ &  0.082 &  0.012 &  0.014 & $+0.008$ & $-0.022$ \\
$ 0.875$ &  0.078 &  0.011 &  0.014 & $+0.007$ & $-0.007$ \\
$ 1.375$ &  0.035 &  0.007 &  0.009 & $+0.005$ & $-0.005$ \\
\hline
\end{tabular}
\caption{\label{xs_kt_di_tab}The cross sections for KTCLUS and
  $\xgo \ge 0.75$.  The third and fourth columns represent the 
  statistical and systematic uncertainties, excluding the principal
  correlated uncertainties which are shown in the fifth column, see
  text.}
\end{center}
\end{table}
\begin{table}
\begin{center}
\begin{tabular}{|r|c|c|c|cc|r@{\hspace{3ex}}|}
\hline
\multicolumn{1}{|c|}{$\bar{\eta}$} & \rule[-3ex]{0ex}{6ex}$\displaystyle\frac{d\sigma}{d\bar{\eta}}$~[nb] & stat.~[nb] & syst.~[nb] & \multicolumn{2}{|c|}{corrl. syst.~[nb]}   \\
\hline
\hline
\multicolumn{6}{|l|}{\rule[-2ex]{0ex}{5ex}$E_T^{min}>6~{\rm GeV}$}\\
\hline
$-0.250$ &   0.56 &   0.05 &   0.20 & $+0.11$ & $-0.08$ \\
$ 0.000$ &   0.89 &   0.06 &   0.24 & $+0.14$ & $-0.14$ \\
$ 0.250$ &   1.35 &   0.07 &   0.33 & $+0.22$ & $-0.16$ \\
$ 0.500$ &   1.47 &   0.07 &   0.19 & $+0.30$ & $-0.26$ \\
$ 0.750$ &   1.49 &   0.08 &   0.36 & $+0.26$ & $-0.21$ \\
$ 1.000$ &   1.16 &   0.07 &   0.15 & $+0.14$ & $-0.16$ \\
$ 1.250$ &   0.57 &   0.05 &   0.14 & $+0.07$ & $-0.07$ \\
$ 1.500$ &   0.26 &   0.03 &   0.07 & $+0.02$ & $-0.04$ \\
\hline
\hline
\multicolumn{6}{|l|}{\rule[-2ex]{0ex}{5ex}$E_T^{min}>8~{\rm GeV}$}\\
\hline
$-0.250$ &   0.09 &   0.02 &   0.08 & $+0.02$ & $-0.02$ \\
$ 0.000$ &   0.25 &   0.03 &   0.04 & $+0.04$ & $-0.04$ \\
$ 0.250$ &   0.37 &   0.03 &   0.06 & $+0.08$ & $-0.06$ \\
$ 0.500$ &   0.45 &   0.04 &   0.09 & $+0.10$ & $-0.07$ \\
$ 0.750$ &   0.59 &   0.05 &   0.25 & $+0.10$ & $-0.08$ \\
$ 1.000$ &   0.55 &   0.05 &   0.07 & $+0.07$ & $-0.09$ \\
$ 1.250$ &   0.31 &   0.03 &   0.09 & $+0.05$ & $-0.04$ \\
$ 1.500$ &   0.23 &   0.03 &   0.05 & $+0.02$ & $-0.04$ \\
\hline
\hline
\multicolumn{6}{|l|}{\rule[-2ex]{0ex}{5ex}$E_T^{min}>11~{\rm GeV}$}\\
\hline
$ 0.375$ &   0.08 &   0.01 &   0.03 & $+0.02$ & $-0.01$ \\
$ 0.875$ &   0.16 &   0.02 &   0.06 & $+0.03$ & $-0.02$ \\
$ 1.375$ &   0.11 &   0.01 &   0.03 & $+0.02$ & $-0.02$ \\
\hline
\hline
\multicolumn{6}{|l|}{\rule[-2ex]{0ex}{5ex}$E_T^{min}>15~{\rm GeV}$}\\
\hline
$ 0.375$ &  0.013 &  0.004 &  0.009 & $+0.003$ & $-0.002$ \\
$ 0.875$ &  0.046 &  0.008 &  0.026 & $+0.007$ & $-0.007$ \\
$ 1.375$ &  0.033 &  0.006 &  0.006 & $+0.002$ & $-0.003$ \\
\hline
\end{tabular}
\caption{\label{xs_kt_re_tab}The cross sections for KTCLUS and
  $0.30<\xgo<0.75$. The third and fourth columns represent the 
  statistical and systematic uncertainties, excluding the principal
  correlated uncertainties which are shown in the fifth column, see
  text.}
\end{center}
\end{table}

\section*{Acknowledgements}

It is a pleasure to acknowledge the efforts of the DESY accelerator
group and the support of the DESY computing group, 
without which this work would not have
been possible. We warmly thank B. Harris, M. Klasen, G. Kramer, 
 and J. Owens for providing
theoretical calculations.

\end{document}